\documentclass{iau}
\usepackage{graphicx}

\title[Asteroseismology, standard candles and the Hubble Constant] 
{Asteroseismology, standard candles and the Hubble Constant:  what is the role of asteroseismology in the era of precision cosmology?}

\author[Neilson et al.]   
{Hilding R. Neilson$^1$, 
  Marek Biesiada$^2$, Nancy Remage Evans$^3$, Marcella Marconi$^4$, Chow-Choong Ngeow$^5$, \& Daniel R. Reese$^6$}

\affiliation{$^1$Dept.~of Physics \& Astronomy, East Tennessee State University, , PO Box 70300, Johnson City, TN 37614, USA  email: {\tt neilsonh@etsu.edu} \\[\affilskip]
$^2$Department of Astrophysics and Cosmology, Institute of Physics, University of Silesia, Uniwersytecka 4, 40-007 Katowice, Poland\\
$^3$Smithsonian Astrophysical Observatory, MS 4, 60 Garden St., Cambridge, MA 02138 \\
$^4$INAF-Osservatorio astronomico di Capodimonte, Via Moiariello 16, I-80131 Napoli, Italy\\
$^5$Graduate Institute of Astronomy, National Central University, Jhong-Li 32001, Taiwan\\
$^6$Institut d'Astrophysique et G{\'e}ophysique de l'Universit{\'e} de Li{\`e}ge,
           All{\'e}e du 6 Aožt 17, 4000 Li{\`e}ge, Belgium

}

\pubyear{2013}
\volume{301}  
\pagerange{???--???}
\jname{Precision Asteroseismology: Celebration of the Scientific Opus of Wojtek Dziembowski}
\editors{W.~Chaplin, J.~Guzik, G.~Hander \& A.~Pigulski, eds.}
\begin{document}

\maketitle

\begin{abstract}
Classical Cepheids form one of the foundations of modern cosmology and the extragalactic distance scale, however, cosmic microwave background observations measure cosmological parameters and indirectly the Hubble Constant, $H_0$, to unparalleled precision.  The coming decade will provide opportunities to measure $H_0$ to $2\%$ uncertainty thanks to the GAIA satellite, JWST, ELTs and other telescopes using Cepheids and other standard candles.  In this work, we discuss the upcoming role for variable stars and asteroseismology in calibrating the distance scale and measuring $H_0$ and what problems exist in understanding these stars that will feedback on these measurements.

\end{abstract}

\firstsection 
\section{Introduction}
The PLANCK satellite, launched May 14th 2009, is another milestone in the cosmic microwave background radiation experiments. It carries an array of 74 detectors aboard, covering the frequency range from 25 to 1000 GHz, and with an angular resolution ranging from $30'$ in low frequency bands to $5'$ in high frequencies. This resolution is a considerable improvement over the previous WMAP experiment. Recently, the PLANCK team released 29 papers summarizing the first year of data analysis. With PLANCK (and with some of its forerunners) we have begun the era of CMB experiments which are more limited by systematics and foregrounds than by the noise. Removing the foregrounds like: thermal and anomalous dust emission, free-free emission, synchrotron emission, cosmic infrared background, CO rotational emission or secondary SZ anisotropies resulted in producing unique maps (for unresolved components) and catalogues (for point sources) of these foreground components. The  PLANCK data are in agreement with a six-parameter $(\Omega_c h^2, \Omega_b h^2, \tau, \theta_{A}, A_s, n_s)$ $\Lambda$CDM model.

However, there are tensions. First of all, the inferred Hubble constant $H_0 = 67 \pm 1.2 \; $km \, s$^{-1} \, $Mpc$^{-1}$ (\cite[Planck Collaboration et al. 2013a]{Planck2013}) is much lower than recently measured on local calibrators: $ H_0 = 73.8 \pm 2.4 \; $km \, s$^{-1} \, $Mpc$^{-1}$ in the SHOES survey  (\cite[Riess et al. (2011)]{Riess2011}) or $H_0 = 74.3 \pm 1.5 \pm 2.1 \; km \, s^{-1} \, Mpc^{-1}$ in the Carnegie Hubble Program  (\cite[Freedman et al. (2012)]{Freedman2012}) based on calibrations of the infrared Leavitt Law (e.g., \cite[Ngeow et al. 2009]{Ngeow2009}; \cite[Freedman et al. 2011]{Freedman2011}). A new measurement, independent of cosmic ladder calibration (\cite[Suyu et al. 2013]{Suyu2013}) obtained from time delays of two strong gravitational lensing systems  RXJ1131-1231 and B1608+656  gave an even higher result: $ H_0 = 75.2 \pm 4.2 \; $km \, s$^{-1} \, $Mpc$^{-1}$. So it's really a sort of tension between PLANCK and other evidence concerning the Hubble constant. There is also another tension in the value of the matter density parameter, $\Omega_m$, which is higher than other independent measurements and even worse it is internally incompatible with PLANCK Sunyaev-Zeldovich data analysis (\cite[Planck Collaboration et al. 2013b]{SZ2013}).

Is this a reason for worry? Every inconsistency is. The strength of modern cosmology lies not in single precise experiments but in consistency across independent and unrelated pieces of evidence. On the other hand it's only the first year of PLANCK mission data which was analyzed -- the rest of the mission awaits the data/results release. At last if we are sure the systematics are all accounted for and tension remains, there could be a signal to go beyond the simplest $\Lambda$CDM model.

Independent local, precise measurements of $H_0$ are needed. 
From a cosmological perspective, in the context of the Dark Energy problem, where the most important goal is to establish details of the expansion and large scale structure formation history, reliably determining $H_0$ in other dedicated experiments would provide invaluable prior information. 
In particular, knowing $H_0$ with $1\%$ accuracy would improve figure of merit in future Dark Energy experiments by $40 \%$ (\cite[Weinberg et al. 2012]{Weinberg2012}). 
\cite[Freedman \& Madore (2010)]{Freedman2010} also noted that at the precision from WMAP observations, one requires better than $2\%$ precision measurements of the Hubble Constant to constrain cosmological parameters. 
 
The cause of the difference between Cepheid and CMB measurements is unclear but in the coming era of GAIA and the James Webb Space Telescope (JWST), it is possible that Cepheids can be employed to measure $H_0$ to better than $2\%$ precision.  When combined with CMB observations, we can expect to also constrain cosmological parameters (e.g., \cite[Planck Collaboration et al. 2013]{Planck2013}).  While GAIA and JWST will allow for unprecedented precision measurements of the Hubble Constant using Cepheids and other standard candles, there are a number of potential challenges as well as opportunities to be considered using precision asteroseismology.

\section{The Role of Cepheids}
\subsection{The Mid-Infrared Cepheid Leavitt Law}
The Cepheid Leavitt Law or PL relation plays an essential role in the extra-galactic distance scale ladder (for reviews of the Cepheid PL relation and its role in distance scale work, see \cite[Madore \& Freedman 1991]{madore1991}; \cite[Ngeow 2012]{ngeow2012}; Ngeow et al. in this Proceeding). The Cepheid PL relation has been well developed in the optical bands, as well as in the near infrared JHK bands. Recently, the calibration of the Cepheid PL relation has moved to the mid infrared (MIR, mainly for the $3.6~\mu\mathrm{m}$- and $4.5~\mu\mathrm{m}$-band) for the LMC Cepheids (\cite[Freedman et al. 2008; Ngeow \& Kanbur 2008; Scowcroft et al. 2011]{freedman2008,ngeow2008,scowcroft2011}), the SMC Cepheids (\cite[Ngeow \& Kanbur 2010]{ngeow2010}), and the Galactic Cepheids (\cite[Marengo et al. 2010; Monson et al. 2012]{Marengo2010a,monson2012}). Advantages of using MIR PL relations in distance scale work include: (a) extinction corrections in MIR are much smaller than in optical bands, which can be safely ignored; (b) the metallicity effect is expected to be minimal in MIR; (c) the intrinsic dispersion of MIR PL relation is 2-3 times smaller than for optical counterparts; and most importantly (d) JWST operated in MIR, is ideal to measure a Hubble constant with $\sim1$\% error from Cepheids observation out to 100~Mpc.

The metallicity dependency of Cepheid MIR PL relation merits further discussion. The slopes of the $3.6~\mu\mathrm{m}$ PL relation based on $\sim10^3$ LMC and SMC Cepheids are almost identical: $-3.25\pm0.01$ (\cite[Ngeow et al. 2009]{Ngeow2009}) vs. $-3.23\pm0.02$ (\cite[Ngeow \& Kanbur 2010]{ngeow2010}), indicating metallicity does not affect the slope in this wavelength. Synthetic $3.6~\mu\mathrm{m}$ PL relations based on pulsation models, however, suggest the opposite (\cite[Ngeow et al. 2012]{ngeow2012a}). More observations of Cepheids in nearby galaxies that span a wide range in metallicity are needed to resolve this issue, similar to analysis done for optical wavelengths (\cite[Bono et al. (2010)]{b10}). The $4.5~\mu\mathrm{m}$ PL relation, on the other hand, is affected by CO absorption (\cite[Marengo et al. 2010]{Marengo2010a}) and could be potentially used as metallicity indicator (\cite[Scowcroft et al. 2011]{scowcroft2011}).



\subsection{Classical Cepheids and metallicity dependence}

In order to predict and interpret the observed pulsation properties of Classical Cepheids and their role as primary distance indicators, nonlinear convective hydrodynamical models are required (e.g.,\cite[Wood et al. 1997; Bono et al. 1999]{bms99,was97}).
In particular, extensive sets of nonlinear convective pulsation models at various chemical compositions have been computed (e.g., \cite[Bono et al. 1999; Fiorentino et al. 2002; Marconi et al. 2005, 2010]{bms99,f02,mmf05,m10}) to provide a theoretical calibration of the
extragalactic distance scale and an accurate investigation of metallicity and helium content effects (\cite[Caputo et al. 2000; Marconi et al. 2005, 2010; Fiorentino et al. 2007; Bono et al. 2008, 2010]{cmm00,f07,mmf05,b08,b10,m10}).
The main result of these investigations is that optical PL relations have significant spread (the intrinsic dispersion is about 0.2-0.3 mag) and depends on the chemical composition of the host galaxy, reflecting the topology and finite width of the instability strip. Moreover, they are affected by nonlinearity at the longest periods or a break around 10 days.  All these systematic effects are reduced when moving toward the near infrared  bands or when introducing the color term in a Period-Luminosity-Colour or Wesenheit relation (e.g., \cite[Caputo et al. 2000; Marconi 2009; Bono et al. 2010]{cmm00,marconi09,b10} and references therein). In particular both observational and theoretical investigation find that the V,(V-I)-Wesenheit relation has a negligible dependence on chemical composition (\cite[Bono et al. 2010; Marconi et al. 2010]{b10,m10}). 

By transforming the theoretical scenario into the HST filters \cite{f13} compared their model predictions with the observed properties of Cepheids in the HST sample galaxies adopted by \cite[Riess et al. (2011)]{Riess2011}, who provided an estimate of the Hubble constant at the 3 \% level of uncertainty.  \cite[Fiorentino et al. (2013)]{f13} concluded that when adopting the predicted PL relation in the HST infrared (160W)  filter the inferred extragalactic distance scale, and corresponding Hubble constant, is in agreement with that obtained by \cite[Riess et al. (2011)]{Riess2011}.
 
\subsection{The Ultra Long Period Cepheids}

The Ultra Long Period Cepheids (ULPs) are fundamental-mode Cepheid-like pulsators with P $\ge$ 80 days (see \cite[Bird et al. 2009]{Bird09} for an extensive discussion), and were first identified in the Magellanic Clouds (\cite[Freedman et al. 1985]{Freedman85}). They are much brighter (M$_{\rm{I}}$ from -7 to -9 mag) than short-period
Cepheids, hence HST is able to observe them up to distances of 100 Mpc.
These pulsators have been identified both in metal rich (\cite[Riess et al. 2009, 2011; Gerke et al. 2011]{riess09,gerke11,Riess2011}) and  metal poor (\cite[Pietrzy{\'n}ski et al. 2004, 2006; Gieren et al. 2004; Fiorentino et al. 2010]{pietr04,pietr06,gieren04,f10}) galaxies.
The two longest-period ULPs have been discovered in IZW18 (\cite[Fiorentino et al. 2010]{f10}), the most metal poor galaxy containing Cepheids, and their behaviour seems to be consistent with the extrapolation of the Cepheid Wesenheit PL Law to higher masses and luminosities (\cite[Marconi et al. 2010]{m10}). Computation of hydrodynamical models of ULP Cepheids is in progress, but new extensive observations allowing the sampling of more than one pulsation cycle are needed in order to confirm their nature. Understanding these pulsators is crucial because, thanks to their brightness, these pulsators are in principle able to reach cosmologically interesting distances in one step (\cite[Fiorentino et al. 2012]{f12}).
Finally we note that the GAIA satellite will provide a direct calibration of
LMC ULPs with parallaxes at the $\mu$arcsec accuracy.

\subsection{Convective Core Overshoot \& Mass Loss}
One of the key issues in understanding classical Cepheids from the perspective of stellar evolution is the Cepheid mass discrepancy (\cite[Keller 2008]{Keller2008}), for which there are two likely solutions.  The first being convective core overshoot during main sequence evolution (\cite[Huang \& Wiegert 1983]{Huang1983}) and the second being Cepheid mass loss (\cite[Neilson \& Lester 2008]{Neilson2008}; \cite[Marengo et al. 2010]{Marengo2010}).  \cite[Neilson et al. (2011)]{Neilson2011} suggested that the mass discrepancy could be solved by a combination of the two possibilities and \cite[Neilson et al. (2012a,b)]{Neilson2012a, Neilson2012b} showed that observed rates of period change provide an observational avenue to distinguish the two phenomena and that mass loss is a ubiquitous property of Galactic Cepheids.  By understanding the role of mass loss in Cepheids, we can better constrain Cepheid properties, i.e., mass-luminosity relation, and how infrared excess from a stellar wind might contaminate the Cepheid Leavitt Law (\cite[Neilson et al. 2010]{Neilson2010}).

\subsection{Cepheid Masses \& Binarity}
Cepheids are of considerable value to the extragalactic distance
scale.  Their value depends on how well we understand them, and that
understanding is linked to how well we know their masses.  Measured 
masses, of course, come from binary systems.  The foundation is the
ground-based orbits of Cepheids in spectroscopic binaries. These lack
the inclination, however there are four ways in which Cepheid masses
are determined in the Milky Way, in addition to the exciting results
for the double-lined eclipsing binaries in the LMC.  
\begin{enumerate}[(1)]
\item  High resolution ultraviolet spectra from satellites (HST and IUE)
have created several double-lined spectroscopic binaries.  Measuring
the orbital velocity amplitude of the hot companion, and combining it
with the orbital velocity amplitude of the Cepheid from the
ground-based orbit and the mass of the main sequence companion from
the spectral energy distribution provides the Cepheid mass (S Mus and
V350 Sgr with HST; SU Cyg with IUE; e.g., \cite[Evans
et al. 2011]{Evans2011} and references therein).
\item  Astrometric motion has been detected in HST FGS observations of
Cepheids (Benedict, et al., 2007).  Combining that with the
spectroscopic orbit of the Cepheid and the mass of the secondary
(\cite[Evans et al. 2009]{Evans2009}) provides the mass of the Cepheid (W Sgr and FF Aql).  
\item  Astrometric motion of both stars in the Polaris system combined
with the spectroscopic orbit provides a fully dynamical mass (\cite[Evans
et al. 2008]{Evans2008}).
\item Interferometry of the V1334 Cyg system (\cite[Gallenne et al. 2013]{Gallenne2013})
has detected the companion and demonstrated orbital motion.  This
approach, particularly when applied to a number of Cepheids promises
to increase the number of measured masses in the Milky Way
significantly.
\end{enumerate}

Continuing work is in progress on improving the accuracy of the masses
and extending the list using interferometric measurements.  Comparison
of masses with evolutionary tracks provided by G.~Bono favors moderate
main sequence core convective overshoot.  

\subsection{Beyond the LMC}
Another avenue to more precisely calibrate the Cepheid Leavitt Law is to use M31 Cepheids as an anchor instead of LMC Cepheids.  The distance modulus of the LMC is about $18.5 \pm 0.1$ (e.g., \cite[Wagner-Kaiser \& Sarajedini 2013]{Wagner2013}; \cite[Dambis et al. 2013]{Dambis2013}; \cite[Marconi et al. 2013]{Marconi2013}). The uncertainty in the LMC distance is one of the largest sources of uncertainty for calibrating the Leavitt Law. An alternative calibration source is M31 Cepheids. The distance to M31 has been measured to $3\%$ precision (\cite[Riess et al. 2012]{Riess2012}; \cite[Valls-Gabaud et al. 2013]{Valls2013}). This increased precision allows for an independent calibration of the Leavitt Law and in the era of JWST, M31 Cepheids will help measure $H_0$ to better precision.

\section{Beyond Classical Cepheids}
 RR Lyrae variable stars are arguably another of the most powerful standard candles (\cite[Caputo 2012]{Caputo2012}; \cite[Marconi 2012]{Marconi2012}; \cite[Cacciari 2013]{Cacciari2013}).  Current calibrations of the RR Lyrae PL relation have allowed for precision measurements of the distance and structure of the LMC (e.g., \cite[Haschke et al. 2012]{Haschke2012}), the structure of the Galactic halo (e.g., \cite[Sesar et al. 2013]{Sesar2013}) and the properties of globular clusters as far as M31 (e.g., \cite[Contreras Ramos et al. 2013]{Contreras2013}). Further, the RR Lyrae PL relation has been calibrated to similar precision as the Cepheid Leavitt Law, especially at infrared wavelengths (e.g.,\cite[C\'{a}ceres \& Catelan 2008]{Caceres2008};  \cite[Klein et al. 2011]{Klein2011}; \cite[Madore et al. 2013]{Madore2013}).  
 
 However, RR Lyrae stars also  suffer from similar uncertainties as Cepheids as standard candles. One such challenge is the lack of measured distances to field RR Lyrae stars; HST astrometric parallax has been measured for only five RR Lyrae stars (\cite[Benedict et al. 2011]{Benedict2011}), meaning most PL relations must be calibrated using distances measured by alternative methods.  There is also an observed metallicity dependence (e.g., \cite[Dambis2013]{Dambis2013}). Furthermore, RR Lyraes also undergo the Blazhko effect (\cite[Buchler \& Koll\'{a}th 2011]{Buchler2011}) and strong photospheric shocks have been observed (\cite[Chadid \& Preston 2013]{Chadid2013}).

If one considers stars with multiple oscillation frequencies, then it is
possible to apply more sophisticated tools from asteroseismology in order to
estimate various stellar parameters, which in turn may lead to more precise
estimates of their luminosities, hence distances. The frequency spectra of
solar-type oscillators, including both main sequence stars such as the Sun and
red giants, follow well-defined patterns.  From these spectra it is possible to
extract $\Delta \nu$, the large frequency separation (i.e. the separation
between modes of consecutive radial order), and $\nu_{\mathrm{max}}$, the
frequency at maximum amplitude.  These quantities then intervene in
scaling relations (\cite[Kjeldsen et al. 1995; Belkacem et al. 2011]{Kjeldsen1995, Belkacem2011})  that have been used to obtain masses and radii of hundreds
of red giants with quoted error bars of 7$\%$ and 3$\%$, respectively (e.g. \cite[Kallinger et al. 2010; Mosser et al. 2010]{Kallinger2010,Mosser2010}).  Such
relations are subject to various uncertainties.  For instance, the outer portion
of solar-type stars are convective and give rise to surface effects which are
poorly modeled. \cite[Kjeldsen et al. (2008)]{Kjeldsen2008} recently proposed a recipe for
correcting/removing such effects from the frequencies.  Another source of
uncertainty is the structural differences between the reference star/model
(typically the Sun) used in a scaling relation and the stars to which the
relation is applied. In order to minimize such effects, one can choose reference
models which are closer to the observed stars, and/or apply seismic inversions. 
Recently, \cite[Reese et al. (2012)]{Reese2012} showed how to directly invert for the mean density
of a star.  Such an approach seeks the frequency combination which is least
sensitive to structural differences between the star and the reference model,
yet minimizes the effects of observational errors.  Applying such an approach
may diminish error bars by 0.5 to 1$\%$ compared to the more traditional scaling
relation between $\Delta \nu$ and the mean density.

Pulsating variable stars and Cepheids will anchor the cosmic distance scale, but there are a number of other standard candles that can be employed to measure $H_0$, including the tip of the red giant branch (\cite[Lee et al. 1993]{Lee1993}, the Tully-Fisher relation (\cite[Tully \& Fisher 1977]{Tully1977})  and Type Ia supernovae (\cite[Phillips 1993]{Phillips1993}).

The tip of the RGB is a powerful standard candle for calibrating primary distance indicators such as Cepheids.  This method is based on the observation that the the tip of the RGB occurs at a constant luminosity and has been calibrated for the LMC (\cite[Salaris \& Girardi 2005]{Salaris2005}) and globular clusters (\cite[Bellazzini et al. 2001]{Bellazzini2001}). The tip of the RGB, however, depends on the age and metallicity of clusters as well as calibrations from either stellar evolution calculations (\cite[Valle et al. 2013]{Valle2013}) or empirical measurements (\cite[Bellazzini et al. 2001]{Bellazzini2001}).  These uncertainties need to be understood in greater detail for the tip of the RGB to be used as a distance indicator to measure $H_0$ to unprecedented precision.

The Tully-Fisher relation is another important standard candle, where the luminosity of a spiral galaxy is related to the rotational velocity of a galaxy.  Because the Tully-Fisher relation can be applied at much greater distances than Cepheids, it is an important complement for measuring $H_0$. Recently, \cite[Sorce et al. (2013)]{Sorce2013} calibrated the relation at $3.6~\mu$m using the same thirteen galaxy clusters as described by \cite[Tully \& Courtois (2012)]{Tully2012}, who calibrated the relation in the $I$-band.  In both works, the authors measured $H_0$ to be approximately $75 ~$km~s$^{-1}$~$M$pc$^{-1}$.  However, the Tully-Fisher relation has a number of uncertainties and corrections, such as a color correction (\cite[Sorce et al. (2013)]{Sorce2013}) and depends on the calibration of the Cepheid Leavitt Law.

Type Ia supernovae were instrumental for the discovery of dark energy  (\cite[Riess et al. 1998]{Riess1998}; \cite[Perlmutter et al. 1999]{Perlmutter1999}) and have been exploited for measuring $H_0$ to unprecedented precision (\cite[Riess et al. 2011]{Riess2011}; \cite[Barone-Nugent et al. 2012]{Barone2012}).  The challenge for employing Type Ia SNe is observing them in nearby galaxies with independently-determined distances.  At large redshift, these standard candles can be calibrated using photometric and spectroscopic redshifts, but in the local group, the calibration requires using Cepheids or other standard candles (\cite[Phillips et al. 2006]{Phillips2006}; \cite[Riess et al. 2011]{Riess2011}).  While Type Ia SNe are one of best standard candles, they are still not ideal.  For instance, the luminosity of Type Ia SNe differ as a function of host galaxy properties (\cite[Sullivan et al. 2011]{Sullivan2010}).

\section{Summary}
The next decade is bright for measuring $H_0$ to less than $2\%$ precision and maybe even $1\%$ thanks to new observatories and satellites such as GAIA, JWST, LSST and the upcoming extremely large telescopes.  The Cepheid PL relation will be calibrated by measuring distances to thousands of Cepheids by GAIA (\cite[Dennefeld, 2011]{Dennefeld2011}), while JWST and LSST will observe Cepheids as far as 100 Mpc (\cite[Freedman \& Madore 2012]{Freedman2012}).  

While the future is bright, there are many current opportunities as well.  Surveys such as VVV, VMC and OGLE are constraining Cepheids and variable star properties and physics.  For instance, the discovery of Cepheids in eclipsing binary systems are constraining the Cepheid mass discrepancy (\cite[Cassisi \& Salaris 2011]{Cassisi2011}; \cite[Neilson \& Langer 2012]{Neilson2012b}; \cite[Prada Moroni et al. 2012]{Prada2012}; \cite[Marconi et al. 2013]{Marconi2013}) as well as the structure of the Leavitt Law (e.g., \cite[Ngeow et al. 2009]{Ngeow2009}).  These observations also allow for constraining the metallicity and the helium abundances of Cepheids, hence constraining  the effects of variations in the chemical composition on the Leavitt Law (\cite[Marconi et al. 2010]{m10}).  We are further able to employ long-term observations of rates of period change to explore Cepheid mass loss and circumstellar media (\cite[Neilson et al. 2012b]{Neilson2012b}).   As such, some key ingredients for better understanding the Cepheid Leavitt Law include the metallicity dependence, convective core overshoot, mass loss and stellar masses.

There are a number of complementary standard candles that can be calibrated by asteroseismic and radial pulsation measurements.  RR Lyrae variable stars form one such distance indicator for measuring distances to galaxies in the Local group such as the LMC and M31, hence calibrating the Cepheid Leavitt law.  The precision of RR Lyrae standard candles will be improved by GAIA parallax measurements as well as better understanding of metallicity dependencies and their stellar evolution from the red giant branch, i.e., the color dispersion.  Asteroseismic observations of red giant stars provide another potential standard candles, in particular seismic inversions measure stellar mean densities, suggesting the potential for calibrating red giants as standard candles.

Because of the significant improvements in the past decade for calibrating standard candles, it is possible to measure $H_0$ to better than $3\%$ precision and it is likely that the measured uncertainty will be reduces to less than $2\%$ in the coming decade.  At that precision, measurements of $H_0$ using standard candles will complement cosmic microwave background determination of cosmological parameters. Furthermore, measuring more precise values of $H_0$ will also require precision understanding of stellar astrophysics.    




\end{document}